\documentclass[preprint,prb,aps]{revtex4}
\begin{document}
\title{ROLE OF CLOSED PATHS IN THE PATH INTEGRAL APPROACH OF STATISTICAL THERMODYNAMICS.} 
\author{J.P. Badiali} 
\address{Universit\'e Pierre et Marie Curie
4 Place Jussieu, 75230 Paris Cedex 05, France \\
Institute for Condensed Matter Physics of National Academy of Sciences of Ukraine,Lviv 79011, Ukraine.\\
jpbadiali@numericable.com}

\begin{abstract}
Thermodynamics is independent of a description at a microscopic level consequently statistical thermodynamics must produce results independent of the coordinate system used to describe the particles and their interactions. In the path integral formalism the equilibrium properties are calculated by using closed paths and an euclidean coordinate system. We show that the calculations on these paths are coordinates independent. In the change of coordinate systems we consider those preserving the physics on which we focus. Recently it has been shown that the path integral formalism can be built from the real motion of particles. We consider the change of coordinates for which the equations of motion are unchanged. Thus we have to deal with the canonical transformations. The Lagrangian is not uniquely defined and a change of coordinates introduces in hamiltonians the partial time derivative of an arbitrary function. We have show that the closed paths does not contain any arbitrary ingredients. This proof is inspired by a method used in gauge theory. Closed paths appear as the keystone on which we may describe the equilibrium states in statistical thermodynamics.

\end{abstract}
\maketitle
\vspace{0.5cm}
 
\section{Introduction}
The path integral method has been proposed by Feynman (\cite{feyn1}) as an alternative to the Schr\"{o}dinger equation. Today it is one theoretical tool among the most used in quantum field theory (\cite{glimm}). The path integral formalism or the functional integral point of view has been extended in statistical thermodynamics (see for instance (\cite{feyn2})). In quantum physics this formalism has been extensively used but due to the presence of a complex measure it is difficult to have rigorous mathematical treatments. In contrast in statistical physics the measure is positive and the functional analysis has a rigorous mathematical basis (\cite{glimm}).\\ 
In the initial version of statistical physics the path integral formalism requires to solve the Sch\"{o}dinger equation, to use the canonical form of the density matrix and to introduce some mathematical tricks. Thus all the machinery of quantum mechanics is needed to calculate thermodynamic quantities. Feynman ((\cite{feyn1}) see '' Remarks on methods of derivation'' p. 295) suggested that it must be possible to calculate more directly the thermodynamic quantities without solving the Schr\"{o}dinger equation but directly starting from the time-dependent motion of particles. In a recent paper (\cite{jpbarxiv}) it has been shown that it is possible to follow this route. The trajectories of particles are quantified by the path integral formalism, only the closed paths have to be inspected on a time associated with a non usual equilibrium condition. In such an approach the time is a real time and the path integral is not just a formal mathematical trick. One interest of this route is to offer a scheme in which thermodynamic equilibrium and irreversible processes can be treated on the same footing (\cite{jpbarxiv}). Schr\"{o}dinger equation and density matrix are not used. It is interesting to note that the thermodynamics of black holes can be obtained without using all the machinery of the Einstein equations (\cite{wald}). 
In this paper we want to show that the thermodynamic properties usually calculated in an euclidean coordinate system are in fact independent of the coordinate system. \\
In Section $2$ we briefly summarize an approach recently proposed in agreement with the Feynman remarks (\cite{jpbarxiv}). In Section $3$ we use a general coordinate system and introduce a traditional result associated with canonical transformations. In Section $4$ we develop an approach reminiscent of the one proposed by Fadeev and Popov (\cite{fadeev}) for gauge theories. A short discussion and conclusion is given in Section $5$. 

\section{Thermodynamic quantities in terms of path integrals}
Feynman (\cite{feyn2}) observed that to the density matrix $\rho$ we may associate a quantity $\rho(u) = \exp \frac{-Hu}{\hbar}$ where $u$ has been redefined to be $\beta \hbar$; $\beta = \frac{1}{k_{B}T}$ in which $T$ is the temperature. This quantity verifies the differential equation 
\begin{equation}
\hbar \frac{\partial \rho(u)}{\partial u}= - H \rho(u)
\label{equafif}
\end{equation}
that is formally a Schr\"{o}dinger equation in which we have made the transformation $it = u$. This remark implies that the partition function can be expressed in term of a path integral as it has been done for the wave function. The partition function $Q$ can be written (\cite{feyn2})
\begin{equation}
Q = \frac{1}{N!}\int dx^{N}(0) \int \mathcal{D} x^{N}(t) exp(- \frac{1}{\hbar} \int ^{\beta \hbar}_{0} H(s) ds )
\label{partfunc}
\end{equation}
where $x^{N}(0)$ represents the set of the positions $x^{i}(0)$ occupied by the $N$ particles at the time $t = 0$ and $x^{N}(t)$ is a similar quantity but associated with the time $t$, $\mathcal{D} x^{N}(t)$ is the path integral measure. The calculation of path integrals has been presented in many textbooks (see for instance ((\cite{schulman}), (\cite{kleinert})). In order to calculate $Q$ we introduce a time-discretization having $\delta t = \frac{\beta \hbar}{n}$ as time step, $n$ goes to infinity and $\delta x$ the difference of position corresponding to $\delta t$. The path integral measure is given by   
\begin{equation}
\mathcal{D} x^{N}(t) = \frac{1}{C} \prod^{N}_{i =1} \prod^{n-1}_{j=1} dx^{i}_{j}
\label{measure}
\end{equation}
in which $C$ is a normalization constant and $x^{i}_{j}$ represents the position of the particle $i$ at the time $ j \delta t$, we have $x^{i}_{0} = x^{i}(0)$ and 
$x^{i}_{n-1} = x^{i}(t -\delta t)$. In (\ref{partfunc}) we must take $x^{i}(0) = x^{i}(\beta \hbar)$ showing that we only consider closed paths, this is associated with the fact that in traditional version of statistical mechanics we only focus on the trace of the density matrix. The Hamiltonian is written $H(s) = K(s) + U(s)$ where $K(s)$ is given by   
\begin{equation}
K(s) = \sum^{N}_{i=1} \frac{1}{2}m (\frac{dx^{i}(s)}{ds})^{2} = \sum^{N}_{i=1} K^{i}(s)  
\label{K}
\end{equation}
and 
\begin{equation}
U(s) =\sum^{N}_{i=1} U^{i}(s)
\label{U}
\end{equation} 
where $U^{i}(s)$ is the total potential acting on the particle $i$ located at the point $x^{i}$ at the time $s$. Written in the form (\ref{partfunc}) the calculation is performed as follows. For each particle $i$ we focus on its position $x^{i}(0)$ fixed at the time $0$ and we consider all the closed paths formed from $x^{i}(0)$ and finally we integrate on all the value of $x^{i}(0)$. \\
Using the relation $F = - k_{B} T \ln Q$ between the free energy $F$ and $Q$ and the thermodynamic relation $F = U - TS$ we obtain in terms of path integral
\begin{equation}
S = k_{B} \ln \frac{1}{N!}\int dx^{N}(0) \int \mathcal{D} x^{N}(t) exp(- \frac{1}{h} \int ^{\beta \hbar}_{0} [H(s) - U] ds ) 
\label{spath}
\end{equation}
showing that the entropy is determined by the fluctuations of the internal energy along the paths these fluctuations being such as $\int ^{\tau}_{0}(H(s)-U) ds \approx h$. Instead of the Boltzmann formula in which we count a number of states here we count a number of paths. \\ 
The path integral formalism developed above corresponds to the initial traditional route. In (\cite{jpbarxiv}) we started on the investigation of a real motion in classical space-time (\cite{jpbarxiv}). We count the closed paths explored during a time $\tau$. If we require that $\tau$ must be such the average of energy counted on the paths is identical to the free energy needed to create the system we find that $\tau = \beta \hbar$. The time $\tau$ is a characteristic of equilibrium in no way it must be considered as the relaxation time characterizing how a system in non-equilibrium relaxes toward an equilibrium state. $\tau$ represents the natural unit of time and it introduces a problem of measurement for the thermodynamic properties (\cite{jpbarxiv}). If a measurement is performed on a time interval smaller than $\tau$ the result will be unpredictable due to the quantum fluctuations. With this point of view 
all the equilibrium properties are calculated from a dynamic point of view but the results are identical to those deduced via the Gibbs ensemble method. \\
To summarize, in this approach the time we consider is the usual time $i.e.$ a real quantity and we have to deal with the usual dynamics. The paths represent the particles trajectories that are quantified via the functional integration in which the Heisenberg uncertainty relations appear. With this approach we do not use the Schr\"{o}dinger equation, this appears as a necessity if we want to use the same formalism for describing both equilibrium states and time-irreversible processes. Another important point of this approach is to show that the passage from statistical mechanics to quantum physics can be analyed in term of time-irreversibility.   

\section{The canonical transformations}
The Hamiltonian $H(s)$ introduced in (\ref{partfunc}) and defined by (\ref{K}) and (\ref{U}) is given in an euclidean coordinate system. The thermodynamic quantities ignoring a microscopic description must be coordinates independent. Thus we have to rewrite $H(s)$ in a general coordinate system. Of course among all the coordinates system we must keep those preserving the investigated physics. We decide to inspect all the coordinate systems verifying the same equation of motion. This leads to focus on the canonical transformations (\cite{arnold}). \\
To save the notations we consider a system reduced to one particle, the generalization is straightforward. The kinetic energy (\ref{K}) is now reduced to $K(x,s) = \frac{1}{2}m (\frac{dx(s)}{ds})^{2}$ and the potential $U(x, s)$ is the external potential. Now we restrict $x$ to represent an euclidean coordinate system and we use the symbol $q$ for a general coordinate system. 
Basically the motion is described via the optimization of a Lagrangian giving rise to the Euler equations. However two lagrangians differing by the total time derivative of an arbitrary function depending on time and position give to the same motion. Let consider two general coordinate systems for which the positions are referred by $q$ and $q'$. Then we have for the change in hamiltonians (\cite{arnold})
\begin{equation}
H'(q',t)) = H(q, t) + \frac{\partial F(q,t)}{\partial t}
\label{G}
\end{equation}
in which $F(q, t)$ is an arbitrary function. Now we have to calculate on the closed paths 
\begin{equation}
\int ^{\tau}_{0} H(q',t) dt = \int ^{\tau}_{0} [H(q, t) + \frac{\partial F(q,t)}{\partial t}]dt 
\label{F1}
\end{equation} 
The integral on $\frac{\partial F(q,t)}{\partial t}$ can be performed easily, we have
\begin{equation}
\int ^{\tau}_{0} \frac{\partial F(q',t)}{\partial t}dt = \delta t \sum ^{n}_{j=1}\frac{F(q_{j},t_{j}) - F(q_{j-1},t_{j-1})}{\delta t} = F(q_{0},\tau) - F(q_{0},0)
\label{F2}
\end{equation}
Thus the integration on closed paths starting from $q_{0}$ introduces to a factor $\Lambda(q_{0}) = \frac{1}{\hbar}[F(q_{0},\tau) - F(q_{0},0)]$. To derive this result we have taken into account that the path integral 
requires a discretization but in this case the result is identical to the one obtained by considering continuous variables.\\
Now we can consider two results obtained with two different coordinates one referred by $q'$ and the second by $q$. We have for (\ref{partfunc}) in the case $N=1$
\begin{equation}
Q' = \int dq'(0) \int \mathcal{D}q'(t) exp(- \frac{1}{\hbar} \int ^{\tau}_{0} H(q',t) dt )
\label{qprime1}
\end{equation}
 \begin{equation}
Q = \int dq(0) \int \mathcal{D}q(t) exp(- \frac{1}{\hbar} \int ^{\tau}_{0} H(q,t) dt )
\label{q1}
\end{equation}
It is well know (\cite{rivers}) that functional measure is invariant in a translation of the variables producing a change of coordinates and hence $dq_{0} \mathcal{D}q(t) = dq'_{0} \mathcal{D}q'(t) $ but with this result and (\ref{F1}) we transform (\ref{qprime1}) as 
\begin{equation}
Q'= \int dq(0) \int \mathcal{D} q(t) exp(- \frac{1}{\hbar} \Lambda(q_{0}) \int ^{\tau}_{0} H(q,t) dt )
\label{qtrois1}
\end{equation}
from which we can see that $Q$ and $Q'$ are different. The closed paths induce a monodromy defined by $\Lambda(q_{0})$. We have to deal with something similar to a gauge theory (see for instance (\cite{rivers})): the basic equations give the same physics but they contain an arbitrary function $F(q,t)$ creating different expressions for the same quantities. Since the physical results must be independent of $F(q,t)$ it must exist a degree of freedom from which the we may eliminate it. Due to the similitude with the gauge theory it seems suited to introduce a method efficient in quantum field theory (\cite{fadeev}). 

\section{Invariance on the coordinate systems}
In standard quantum mechanics the expectation $\left\langle A \right\rangle$ of a given operator $A$ is calculated via the density matrix $\rho$ according to  
$\left\langle A \right\rangle = Tr(\rho A)$ where $Tr(B)$ means that we have to take the trace of the operator $B$. To calculate $Tr(\rho A)$ we need to introduce a complete basis of orthogonal and normalized vectors. It is very well known that the trace is independent of the basis used, to demonstrate this we use the closure relation $ \sum_{i}\left|i > < i\right| = 1$ that we can consider as a decomposition of the unity. In addition, in the demonstration of the trace independence we perform a change in the order of the summation. Similar arguments will be used to show the independence of the thermodynamic variables on the choice of the coordinate system. However these arguments are implemented by a method used in gauge theory (\cite{fadeev}).\\
Here the decomposition of the unity wil be build via the Dirac distribution defined according to 
\begin{equation}
\int\delta (y - x) f(x)dx = f(y)
\label{dirac1}
\end{equation}
For each value of $q_{0}$ we have associated a number $\Lambda(q_{0})$ depending on the arbitrary function $F(q,t)$. By changing, for instance, some parameters involved in 
the definition of $F(q,t)$ we can change its value and we assume that it can reach a value $\kappa (q_{0})$ where $\kappa (q_{0})$ is a given function of $q_{0}$. Using (\ref{dirac1}) for the function $f(x) =1$ we get
\begin{equation}
\int \delta (\kappa (q_{0}) - \Lambda(q_{0})) d\Lambda(q_{0}) = 1
\label{dirac2}
\end{equation}
If (\ref{dirac2}) is inserted in (\ref{qtrois1}) the result is unchanged and independent of $\kappa (q_{0})$ and we can perform an integration on a normalized function $g(\kappa (q_{0}))$. Now we have
\begin{equation}
Q'= \int dq_{0 }d\kappa (q_{0}) g(\kappa (q_{0})) d\Lambda(q_{0}) \delta (\kappa (q_{0}) - \Lambda(q_{0})) \exp{ -\frac{1}{\hbar} \Lambda(q_{0})} \int \mathcal{D} q(t) exp(-  \int ^{\tau}_{0} H(q,t) dt )
\label{qtrois2}
\end{equation}  
If we perform first an integration over $\kappa (q_{0})$ we get 
\begin{equation}
Q'= Cte \int dq_{0 }\int \mathcal{D} q(t) exp(-  \int ^{\tau}_{0} H(q,t) dt )
\label{qint1}
\end{equation}
in which the constant $Cte$ is given by 
\begin{equation}
Cte = \int d\Lambda (q_{0}) g(\Lambda (q_{0}))\exp{ -\frac{1}{\hbar} \Lambda(q_{0})}
\label{C}
\end{equation}
The function $g(\kappa (q_{0}))$ which indicates the distribution of $\kappa (q_{0})$ is chosen independently of 
$\Lambda (q_{0})$ and consequently the constant $Cte$ is independent of $\Lambda (q_{0})$. For instance, to be illustrative we can choose for $g(x)$ a Gaussian distribution  
\begin{equation}
g(x) = (\frac{A}{\pi})^{\frac{1}{2}} \exp{A x^{2}}
\label{gauss}
\end{equation}
The value of $Cte$ is $\exp{\frac{1}{4 \hbar^{2}A}}$. Since the dimension of $\Lambda (q_{0})$ is the same as $\hbar$ we can take $A = \frac{1}{\hbar^{2}}$ and finally $Cte$ is just a numerical factor $C = \exp{\frac{1}{4}}$.\\

\section{Discussion - Conclusion}
The use of generalized coordinates has been previously discussed by Dirac (\cite{dirac}) in a formalism connected with the Sch\"{o}dinger equation. He noted that the hamiltonian contains arbitrary functions that does not change the state of the system $i.e.$ the dynamics. In this work we do not use the Schr\"{o}dinger equation but the path integral formalism for describing the thermodynamic properties. Note that in the loop quantum gravity it is important to have results that are background independent, this is realized if also we consider closed paths or loops (\cite{smolin}).\\
What happens if we consider open paths joining two different points in space. This is equivalent to investigate the correlation between particles located at these two points. These correlations are important in order to characterize a system but this is beyond the scope of thermodynamics in which we ignore the existence of particles. For open paths the quantity $\Lambda(q_{0})$ has to be replaced by a function depending on two points in space and we cannot eliminate this function from general arguments as it was possible to do here.\\    
We can say that the exact definition of the partition function is given by (\ref{qint1}), it differs from the traditional one by a numerical factor. But the main result is that the thermodynamic properties are independent of the coordinate system. This is the consequence of the liberty we have to choose the function $g(x)$ which is independent of the arbitrary function $F(q,t)$. One choice of $g(x)$ has been proposed to be illustrative. Our demonstration is inspired by the properties of the trace of the density matrix and by the Fadeev-Popov method (\cite{fadeev}) used in gauge theory. Finally, we may conclude that the closed paths are strongly associated with equilibrium thermodynamics.\\


\end{document}